\documentclass{moriond}
\usepackage{lineno}

\def\be{\begin{equation}}
\def\ee{\end{equation}}
\def\bea{\begin{eqnarray}}
\def\eea{\end{eqnarray}}

\newcommand{\Photo}{}
\input{belle2-symbols}
\begin{document}
\vspace*{4cm}
\title{Hadronic $B$ decays and charm at Belle II}

\author{ Sagar Hazra, on behalf of the Belle II collaboration }

\address{Tata Institute of Fundamental Research,\\
Mumbai 400 005, India}

\maketitle\abstracts{
We report the measurements of various hadronic $B$ decays at the Belle II experiment using a $362\invfb$ sample of electron-positron collisions collected at the $\FourS$ resonance.
All results agree with the previous determination, and some of them are already competitive with the world's best measurement.
In addition, we present a newly developed algorithm for $D$ meson flavor tagging (discriminating between $D^{0}$ and $\bar{D^{0}}$) at Belle II.
}

\section{Introduction}

Measurement of hadronic $B$ decays play an important role in the flavor physics program to test the standard model (SM) and its extensions.
Decays mediated by Cabbibo-suppressed $b\to u$ and $b\to d,s$ loop transitions constitute sensitive probes for non-SM contributions.
We can exploit isospin symmetry in some hadronic decays to construct various sum rules.
One such sum-rule combines the branching fractions and $CP$ asymmetries of $B\to K\pi$ decays, providing a null test with precision better than $1\%$ in the SM~\cite{sumrule}.
Similarly, the CKM angle $\phi_{2}/\alpha$ can be determined by measuring various $B\to\pi\pi$ decays related by isospin symmetry.
Belle II has a unique capability of studying jointly, and within a consistent experimental environment, all relevant final states of isospin-related $B$ decays to put a stringent bound on the sum-rule test  as well as to improve our knowledge of angle $\phi_{2}$ .
The CKM angle $\phi_{3}/\gamma$ is the SM candle for $\CP$ violation and is very reliably predicted.
A measurement of this angle has been performed in an analysis of the $B \to D K$ decays.
Lastly, we report a novel charm flavor tagger that would be important for $\CP$ violation and mixing studies in the charm sector.

\section{Determination of signal yield}

A key challenge in  reconstruction of decay modes considered here is the large contamination from $e^{+}e^{-}\to\qqbar$ $(q=u,d,s,c)$ background coupled with a small signal branching fraction.
We use a binary-decision-tree classifier that combines a number of variables, most related to the event shape topology, which provide discrimination between the $B\bar{B}$ and $\qqbar$ events.
To determine the signal yield, we rely on two kinematic variables: the energy difference $\Delta E = E^{*}_{B}- \sqrt{s}/2$ between the energy of the reconstructed $B$ candidate and half the collision energy, and the beam-energy-constrained mass $\mbc = \sqrt{s/(4c^{4}) -(p^{*}_{B}/c)^2}$,
which is the invariant mass of the $B$ meson, with its energy being replaced by half the collision energy; all quantities are calculated in the $\FourS$ frame.

\section{Isospin sum-rule}

The isospin sum-rule relation for the $B\to K \pi$ system provides a stringent null test of the SM~\cite{sumrule},
\begin{equation}
\label{eq:ikpi}
    I_{K\pi} = \mathcal{A}_{K^+\pi^-} + \mathcal{A}_{K^0\pi^+}\frac{\mathcal{B}(K^0\pi^+)}{\mathcal{B}(K^+\pi^-)}\frac{\tau_{B^0}}{\tau_{B^+}} - 2\mathcal{A}_{K^+\pi^0}\frac{\mathcal{B}(K^+\pi^0)}{\mathcal{B}(K^+\pi^-)}\frac{\tau_{B^0}}{\tau_{B^+}} - 2\mathcal{A}_{K^0\pi^0}\frac{\mathcal{B}(K^0\pi^0)}{\mathcal{B}(K^+\pi^-)}=0,
\end{equation}
where $\mathcal{B}$, $\mathcal{A}$, and $\tau$ are the branching fractions, direct $\CP$ asymmetries, and lifetimes of $B$ mesons, respectively.
We measure the time-integrated asymmetry for the $\CP$ eigenstate $\Bz\to\Kz\piz$ by inferring the $B$-meson flavor ($\Bz$ or $\Bzb$) from that of the other $B$ meson produced on the $\FourS$ decay, using a category-based flavor tagger~\cite{flavortagger}. 

\begin{figure}[htb!]
    \centering
    \includegraphics[scale=0.20]{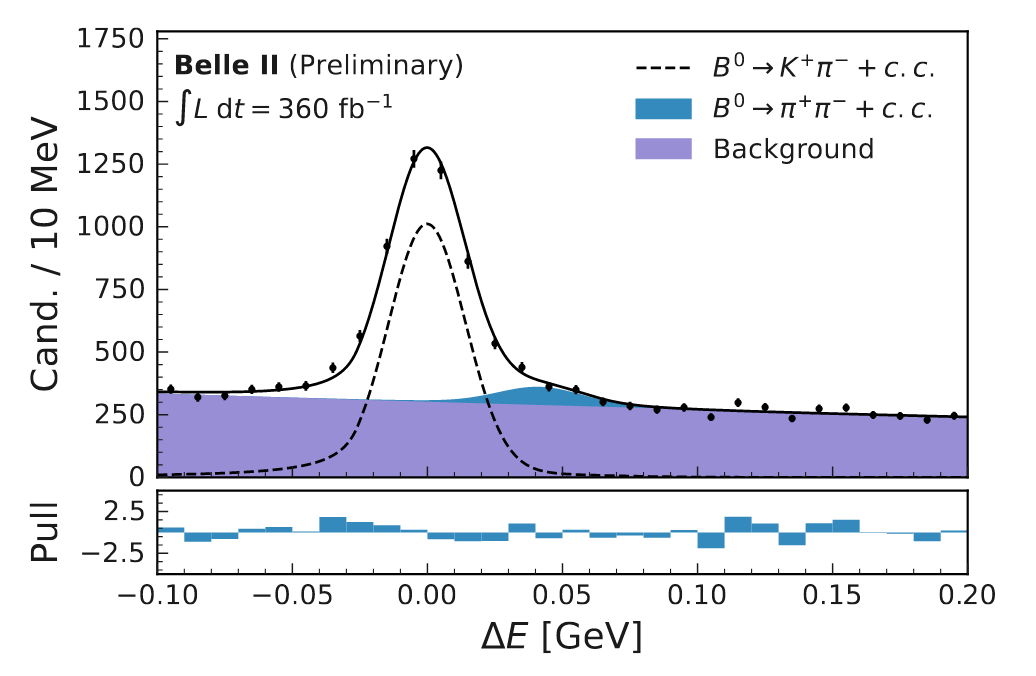}
   \includegraphics[scale=0.20]{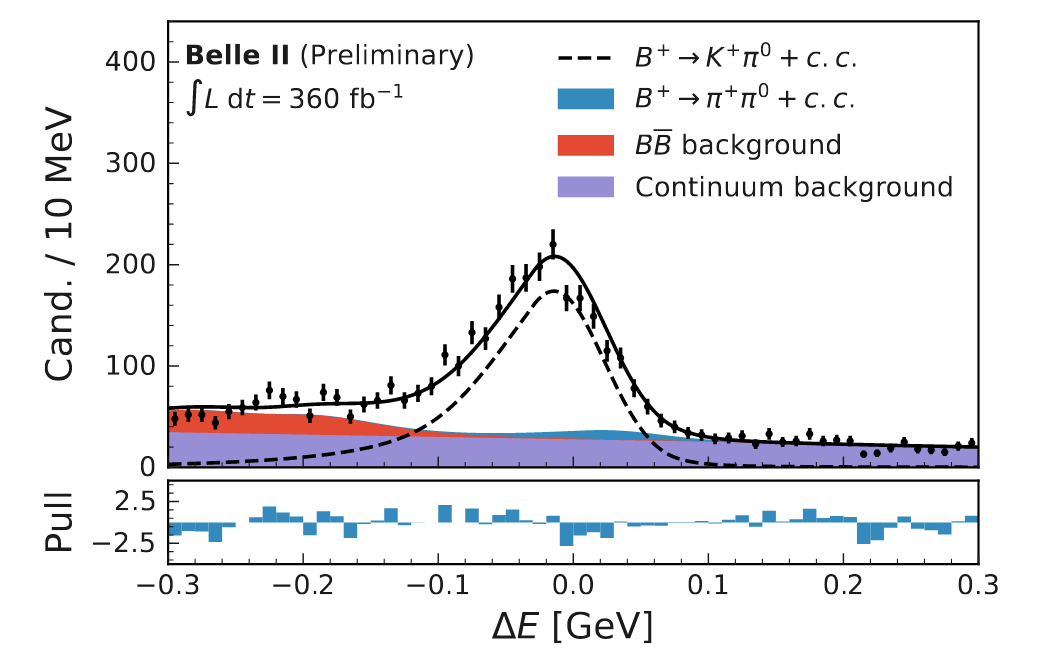} 
    \caption{Signal-enhanced $\Delta E$ distributions of $\Bz \to K^{+} \pi^{-}$ (left) and $\Bp \to K^{+} \pi^{0}$ (right).}
    \label{fig:isospin1}
\end{figure}

\begin{figure}[htb!]
    \centering
    \includegraphics[scale=0.18]{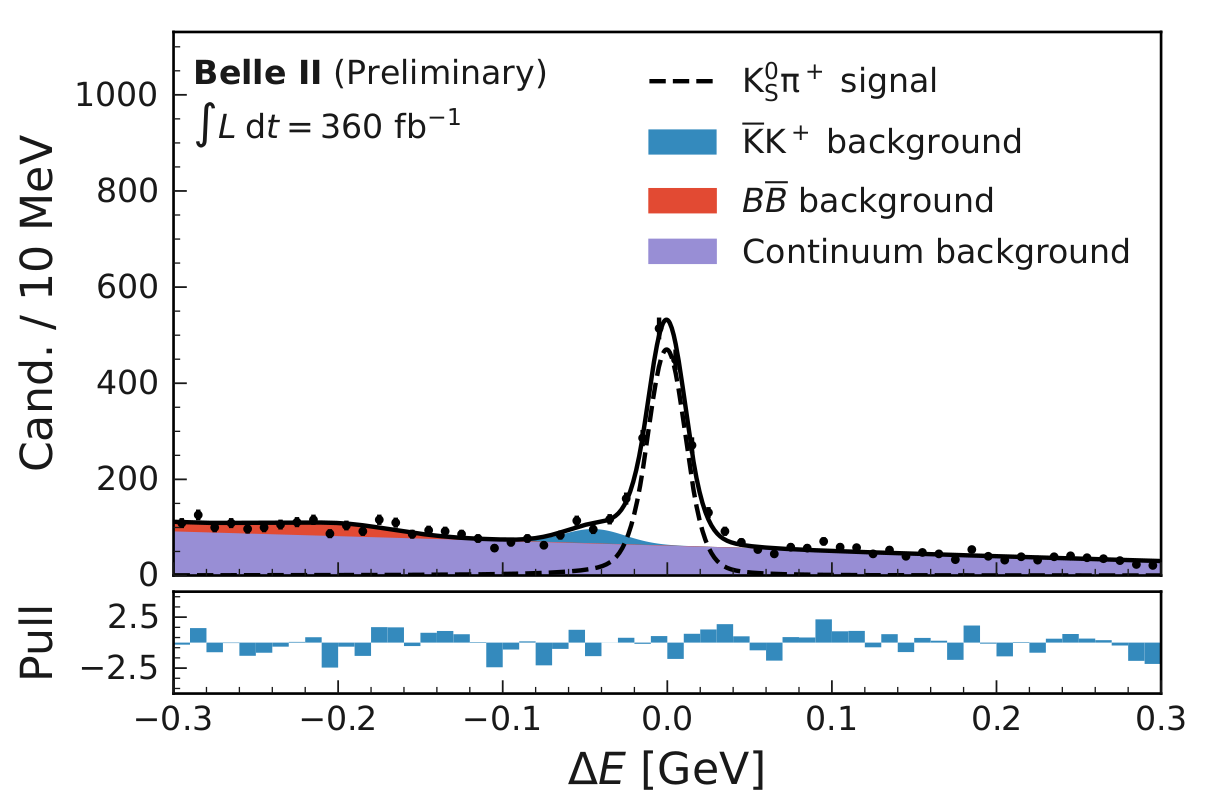}
   \includegraphics[scale=0.18]{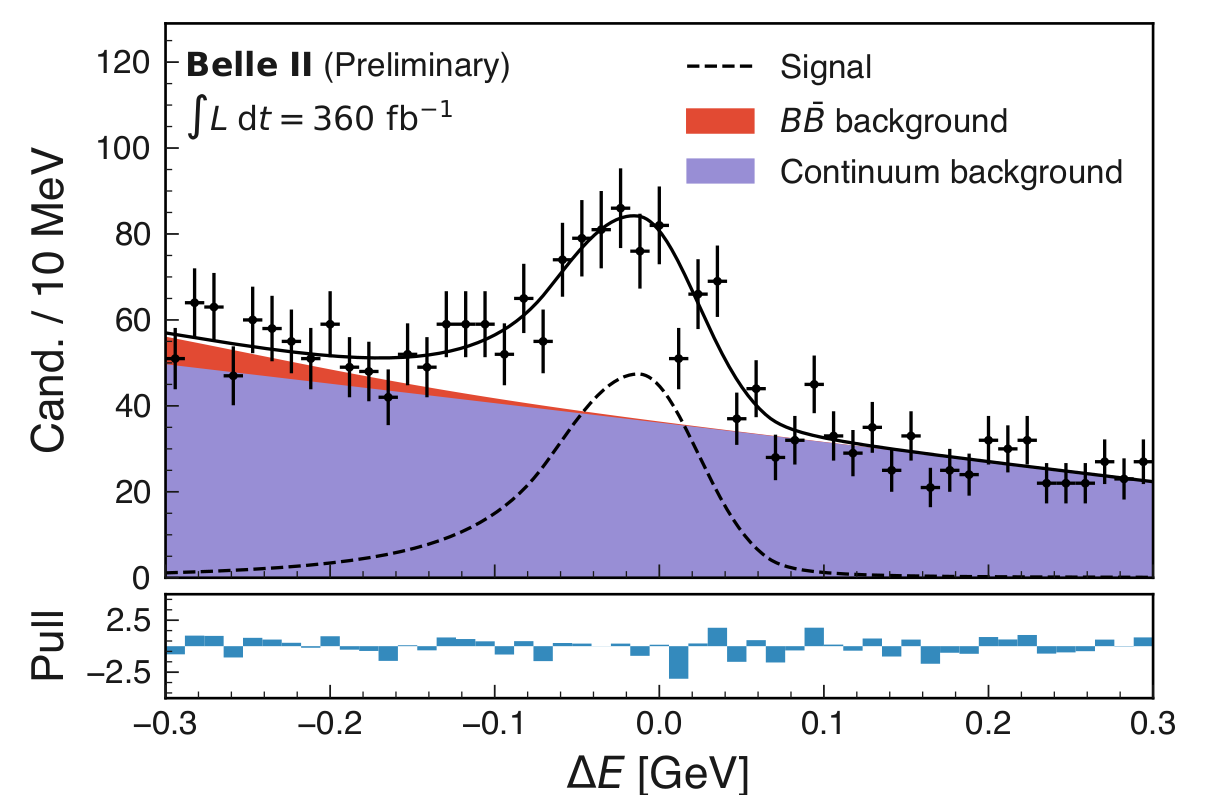} 
    \caption{Signal-enhanced $\Delta E$ distributions of $\Bp \to K^{0} \pi^{+}$ (left) and $\Bz \to \Kz \pi^{0}$ (right).}
    \label{fig:isospin2}
\end{figure}

Figures~\ref{fig:isospin1} and \ref{fig:isospin2} show the $\Delta E$ distributions of all  four $K\pi$ final states.
From the fits we obtain the following branching fractions,
\begin{eqnarray*}
{\mathcal B}(B^{0}\to K^{+} \pi^{-}) &=& [20.7 \pm 0.4\stat \pm 0.6\syst] \times 10^{-6},\\
{\mathcal B}(B^{+}\to K^{+} \pi^{0}) &=& [14.2 \pm 0.4\stat \pm 0.9\syst] \times 10^{-6},\\
{\mathcal B}(B^{+}\to K^{0} \pi^{+}) &=& [24.4\pm 0.7\stat \pm 0.9\syst] \times 10^{-6},\\
{\mathcal B}(B^{0}\to K^{0} \pi^{0}) &=& [10.2\pm 0.6\stat \pm 0.6\syst] \times 10^{-6}
\end{eqnarray*}
and $\CP$ asymmetries
\begin{eqnarray*}
    \ACP (B^{0}\to K^{+} \pi^{-}) &=& -0.07 \pm 0.02 \stat \pm 0.01\syst,\\
    \ACP (B^{+}\to K^{+} \pi^{0}) &=& 0.01 \pm 0.03\stat \pm 0.01\syst,\\
    \ACP (B^{+}\to K^{0} \pi^{+}) &=& -0.01 \pm 0.08\stat \pm 0.05\syst,\\
    \ACP(B^{0}\to K^{0} \pi^{0}) &=& -0.06\pm 0.15\stat \pm 0.05\syst.
\end{eqnarray*}
The dominant contribution to the systematic uncertainties comes from the $\piz$ and $\KS$ reconstruction efficiencies for the decays having these final state particles.
These are determined  with the help of selected control samples and are expected to significantly decrease with the availability of larger sample sizes.

\begin{figure}[htb!]
    \centering
    \includegraphics[scale=0.39]{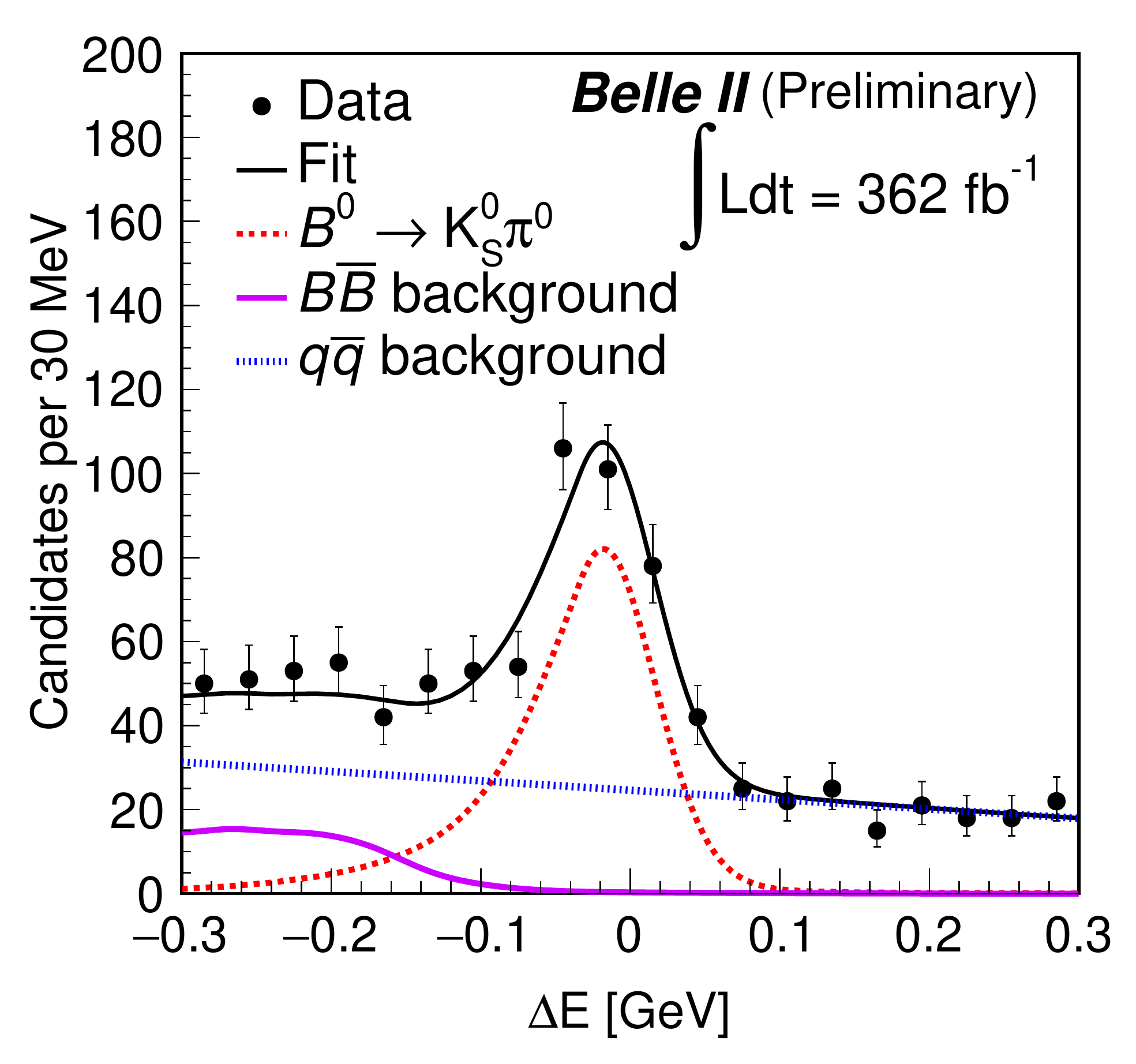}
   \includegraphics[scale=0.39]{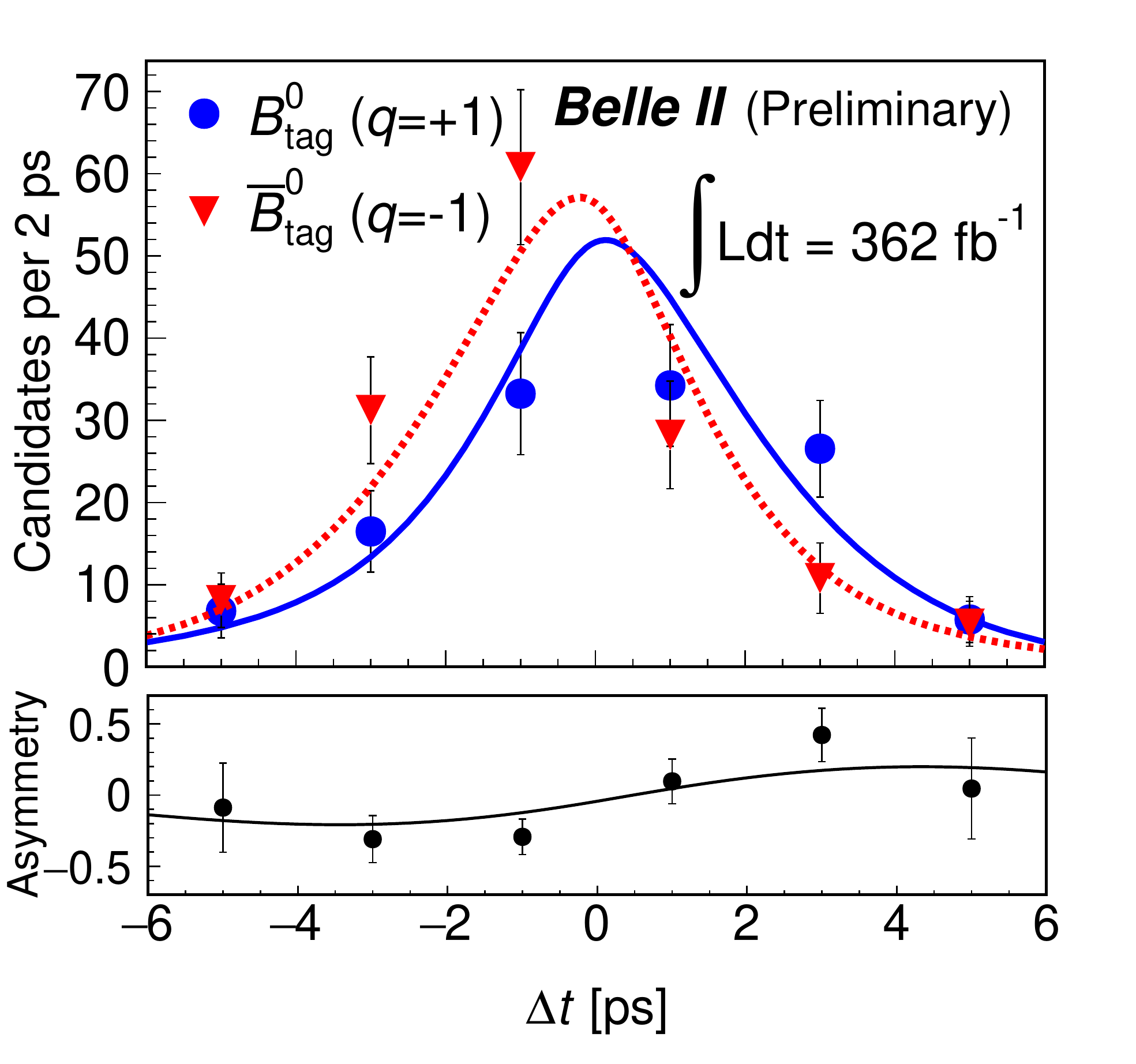} 
    \caption{Signal-enhanced $\Delta E$ distribution (left) and background subtracted $\Bz$ and $\Bzb$--tag $\Delta t$ distribution (right) for $\Bz\to \KS \piz$ time-dependent $\CP$ asymmetry measurement.}
\label{fig:isospin3}
\end{figure}

We also measure $\CP$ asymmetry in $\Bz\to \KS \piz$ decays using a time-dependent method.
Additional motivation to perform this measurement is to determine the value of $\Delta \SCP= \SCP - \sin(2 \phi_{1})$ for the $b\to s$ loop transition, which is sensitive to potential NP contribution.
The main challenge of this analysis is the absence of primary charged particles, which leads to poor decay time resolution.
The analysis is validated with the $\Bz\to J/\psi \KS$ control sample, with the $B$ decay time  reconstructed using only the $\KS$ vertex.
Figure~\ref{fig:isospin3} shows the  reconstructed $\Delta E$ and $\Delta t$ (difference in proper times between two $B$ meson decays) distributions from which we obtain
\begin{eqnarray*}
\ACP=&0.04^{+0.15}_{-0.14}\stat \pm 0.05 \syst
\end{eqnarray*}
and
\begin{eqnarray*}
\SCP = &0.75^{+0.20}_{-0.23} \stat \pm 0.04 \syst.
\end{eqnarray*}
Precision of the measured mixing-induced asymmetry parameter $\SCP$ is already competitive with the world's best measurement  although based on a small dataset.

We combine the time-dependent and time-integrated measurements to obtain the best sensitivity of $\mathcal{A}_{\KS \piz}=-0.01 \pm 0.12 \stat \pm  0.05 \syst$.
Putting all $\mathcal{B}$ and $\ACP$ values of the $K\pi$ system together, we obtain an overall Belle II isospin test:
\begin{eqnarray*}
    I_{K\pi}= -0.03 \pm 0.13 \stat \pm  0.05 \syst, 
\end{eqnarray*}
which is consistent with the SM prediction and comparable with world's best result ($-0.13 \pm 0.11$) even with a smaller sample.

\section{ Towards the determination of {\boldmath $\phi_{2}$}}
\begin{figure}[htb!]
    \centering \includegraphics[scale=0.23]{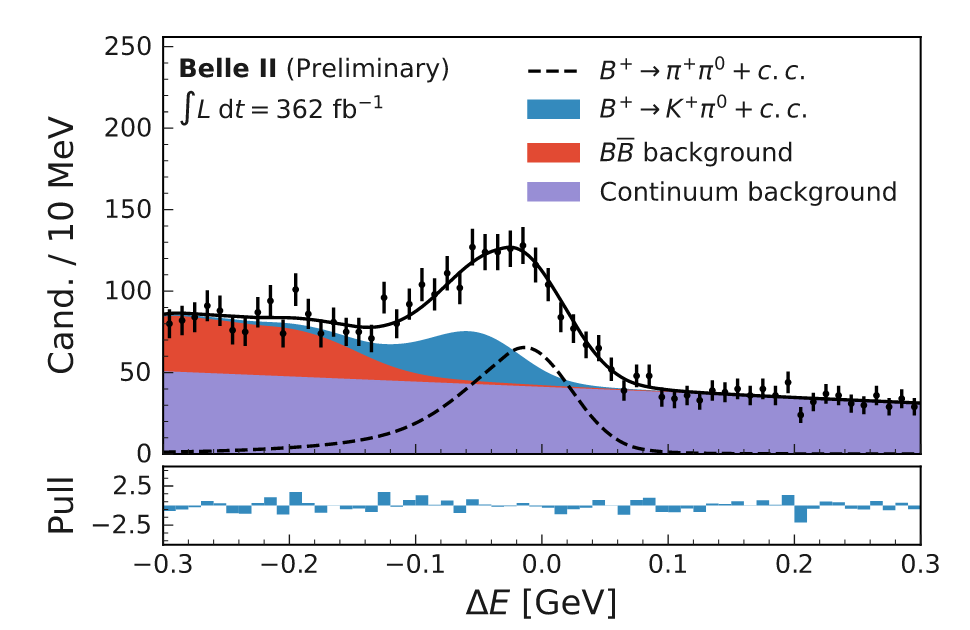}
    \includegraphics[scale=0.20]{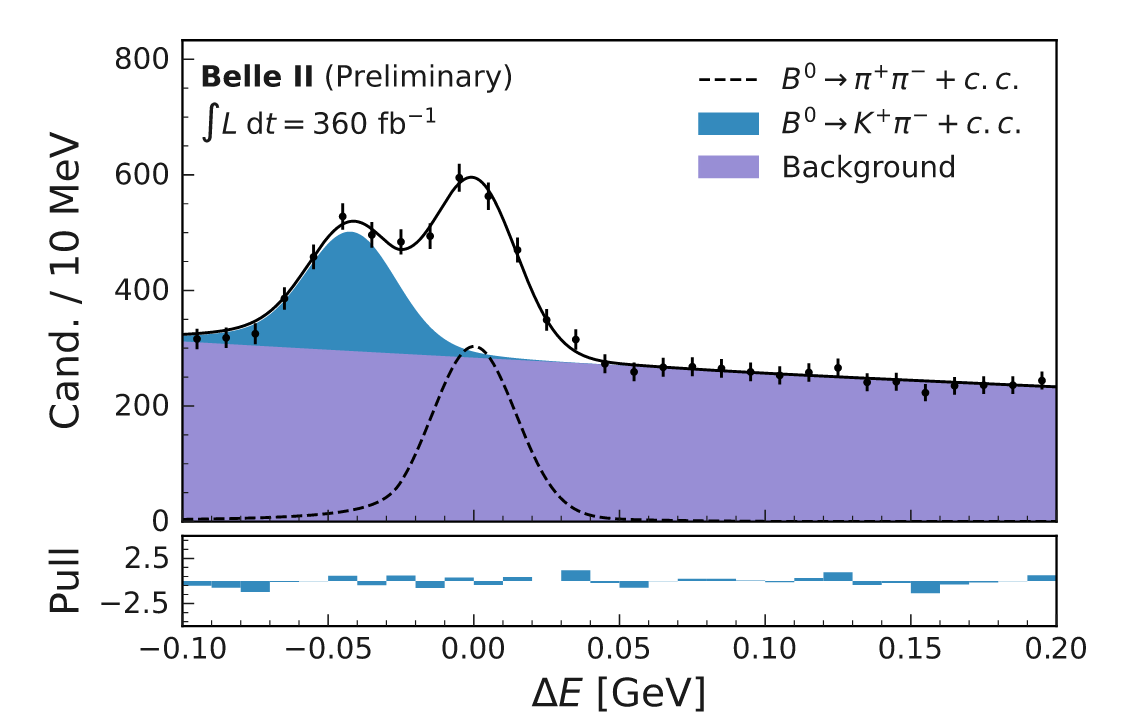}
    \caption{Signal-enhanced $\Delta E$ distributions of $B^{+} \to \pi^{+} \piz$ (left) and $\Bz \to \pi^{+} \pi^{-}$ (right).}
    \label{fig:alpha1}
\end{figure}

  The combined analysis of branching fractions and $\CP$ violating asymmetries of the complete set of $B \to \pi \pi$ isospin partners enables a determination of $\phi_{2}$~\cite{alpha}.
We focus here on $\Bp \to \pip \piz$ and $\Bz \to \pip \pim$ decays.
Belle II has the unique capability to study all the $B\to \pi \pi $ decays to determine the CKM angle $\phi_{2}$.
Figure~\ref{fig:alpha1} shows the $\Delta E$ distributions of $\pi^{+} \piz$ and $pi^{+}\pi^{-}$ channels.

We obtain the following branching fractions,
\begin{eqnarray*}
\mathcal{B}(\Bz \to \pi^{+} \pi^{-}) &=& [5.83\pm 0.22\stat \pm 0.17\syst] \times 10^{-6},\\
\mathcal{B}(B^{+}\to \pi^{+} \pi^{0}) &=& [5.02\pm 0.28 \stat \pm 0.32\syst] \times 10^{-6},
\end{eqnarray*}
and $\CP$ asymmetry of $\ACP (B^{+}\to \pi^{+} \pi^{0})=-0.08 \pm 0.05\stat \pm 0.01\syst$.
The dominant contribution in the systematic uncertainties comes from $\piz$  reconstruction  and tracking efficiency.

\section{Determination of {\boldmath $\phi_{3}/\gamma$}}
The CKM  angle $\phi_{3}/\gamma$ is a SM benchmark as it is the only angle  accessed using  tre
e level $B$ decays.
The angle $\phi_{3}$ is governed by interference between the favoured $b\to c \bar{u} s$ and suppressed $b\to u \bar{c} s$ transitions in the $B\to D K$ decays:
\begin{eqnarray}
\frac{\mathcal{A}_{\rm sup}(B^{-}\to \bar{D^{0}} K^{-})}{\mathcal{A}_{\rm fav} (B^{-}\to \bar{D^{0}} K^{-})}=r_{B} e^{i(\delta_{B}-\gamma)},
\end{eqnarray}
where $\delta_{B}$ is the strong phase difference and $r_{B}$ is the magnitude of the suppression.
The angle $\phi_{3}$  can be measured using different modes based on a different possible $D$ final states.
We present the determination of $\phi_{3}$ using  GLW~\cite{GLW1,GLW2} and GLS~\cite{GLS} methods with Belle and Belle II datasets.

The GLW method uses the $D\to K^{+} K^{-}$ ($\CP$-even) and $D\to \KS \piz$ ($\CP$-odd) eigenstate to determine $\phi_{3}$ from  $\mathcal{R}_{\CP \pm}= 1 + r_{B}^{2} \pm 2 r_{B} \cos \delta_{B} \cos \phi_{3}$ and $\mathcal{A}_{\CP \pm}= \pm 2 r_{B} \sin \delta_{B} \sin \phi_{3}/\mathcal{R}_{\CP \pm}$.
This analysis used a combined Belle ($711 \invfb$) and Belle II ($189 \invfb$)  data sample.
We find the following relative branching frations,
\begin{eqnarray*}
 \mathcal{R}_{\CP+}=(1.16 \pm 0.08 \stat \pm 0.04 \syst)\%,\\
 \mathcal{R}_{\CP-}=(1.15 \pm 0.07 \stat \pm 0.02 \syst)\%
\end{eqnarray*}
and $\CP$-violating rate asymmetries,
\begin{eqnarray*}
\mathcal{A}_{\CP+}=(+12.5 \pm 5.8 \stat \pm 1.4 \syst)\%,\\
\mathcal{A}_{\CP-}=(-16.7 \pm 5.7 \stat \pm 0.6 \syst)\%.
\end{eqnarray*}
While the results for $\CP$-even eigenstate are not yet competitive with the world average, the $\CP$-odd eigenstate results achieve world's best measurement as it is a unique channel for the Belle II.

The GLS method uses the Cabibbo-suppressed channels $B^{\pm}\to D (\to\KS K^{\pm}\pi^{\mp}) h^{\pm}$ (same sign) and
$B^{\mp}\to D (\to\KS K^{\pm}\pi^{\mp}) h^{\mp}$ (opposite sign) to determine 4 $\CP$ asymmetries and 3 branching ratios.
This analysis used the combined Belle ($711 \invfb$) and Belle II ($362 \invfb$) data sample.
While the results are not competitive with world average, they still provide a constraint on the measurement on $\phi_{3}$.
This results will be used for the combination of $\phi_{3}$ measurement with Belle and Belle II data sample. 
We find the following ratio of branching fractions,
\begin{eqnarray*}
\mathcal{A}^{DK}_{SS}=-0.089 \pm 0.091 \pm 0.011,\\
\mathcal{A}^{DK}_{OS}=+0.109 \pm 0.133 \pm 0.013,\\
\mathcal{A}^{D\pi}_{SS}=+0.018 \pm 0.026 \pm 0.009,\\
\mathcal{A}^{D\pi}_{OS}=-0.028 \pm 0.031 \pm 0.009,
\end{eqnarray*}
and $\CP$-violating rate asymmetries,
\begin{eqnarray*}
\mathcal{R}^{DK/D\pi}_{SS}=0.122 \pm 0.012 \pm 0.004, \\
\mathcal{R}^{DK/D\pi}_{OS}=0.093 \pm 0.013 \pm 0.003,\\
\mathcal{R}^{D\pi}_{SS/OS}=1.428 \pm 0.057 \pm 0.002.
\end{eqnarray*}

\section{The charm flavor tagger}

\begin{figure}[htb!]
    \centering \includegraphics[scale=0.27]{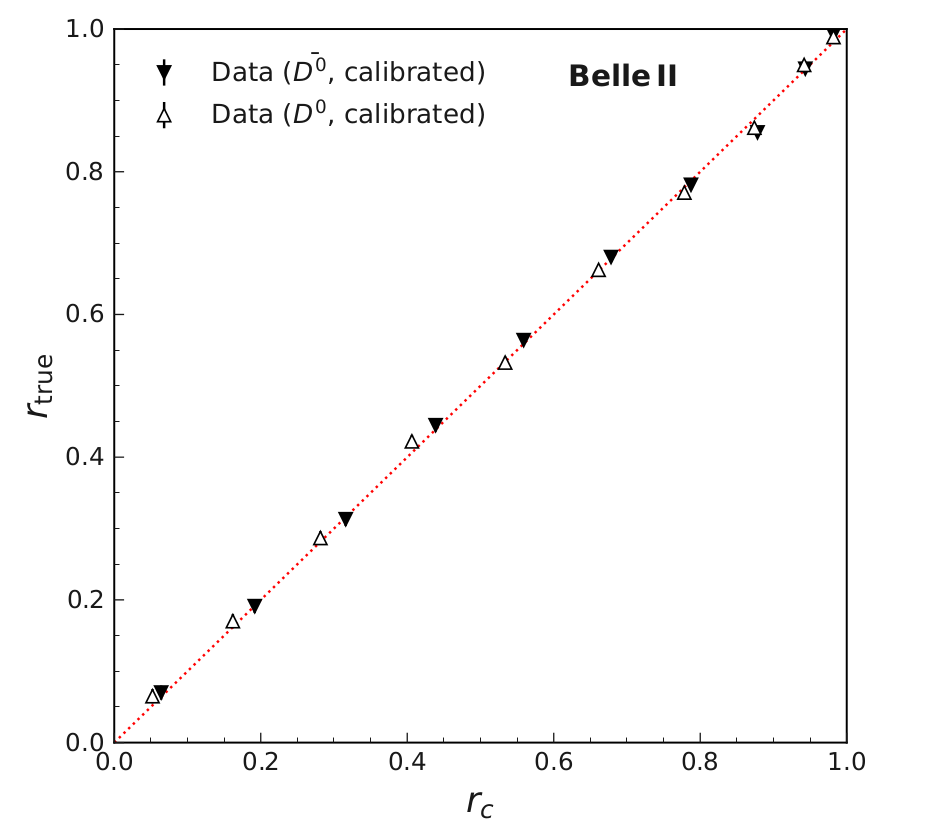}
    \caption{True dilution as a function of calibrated dilution for $D^{0}\to K^{-}\pi^{+}$ decays.}
    \label{fig:calibration}
\end{figure}

\begin{figure}
    \centering
    \includegraphics[scale=0.2]{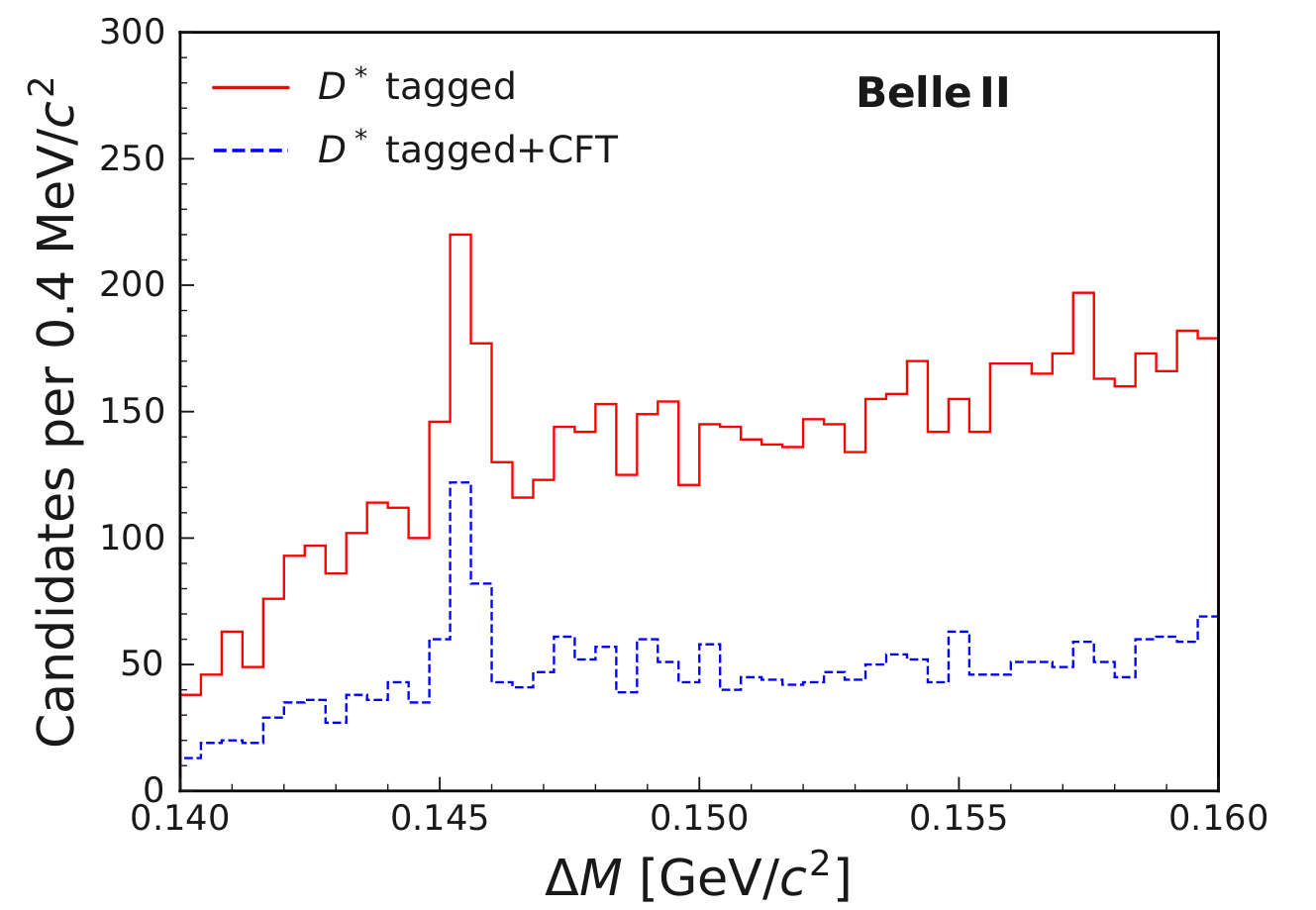}
    \caption{Distribution of the difference between $D^{*}$ and $D^{0}$ mass for the $D^{*}\to D^{0}[\to K^{+}\pi^{-}\piz]\pi^{+}$ decays. }
    \label{fig:CFT}
\end{figure}

Identification of the $\Dz$ flavor plays a crucial role in the $\CP$-violation and mixing measurement in the charm sector.
Typically all the charm analysis uses the conventional $D^{*}$-tagging method which has high purity but substantially reduces the data sample size.
The main motivation for developing a new algorithm is to also include $D^{0}$ mesons that do not emerge from a $D^{*}$ decay.
The new charm flavor tagger uses boosted-decision-trees to recover additional flavor information from the extra charged particles.
Figure~\ref{fig:calibration} shows a good agreement between the calibrated and true flavor dilution.
The novel charm flavor tagger has an effective tagging power,
\begin{eqnarray*}
 \epsilon_{\rm tag}^{\rm eff}=(47.91\pm 0.07\stat \pm 0.51 \syst)\%,
\end{eqnarray*}
which is calculated in the $D^{0}\to K^{-} \pi^{+}$ decays.
Effective increase in the sample size is estimated to evaluate the impact of charm flavor tagger in physics analysis.
Figures~\ref{fig:CFT} shows the effect of charm flavor tagger on $D^{*}\to\Dz[\to K^{+} \pi^{-} \piz] \pi^{+}$ decays.
We find for $\Dz\to K^{-} \pi^{+}$, doubling  the effective sample size compared to conventional $D^{*}$-tagged decays.

\section{Conclusions}

In summary, hadronic $B$ decays and charm physics play an important role in sharpening flavor picture.
Belle II has unique access to channels that offer key tests of the SM.
We have shown five new results:  
$\CP$ violation in $\Bz\to\KS\piz$ that probes isospin sum rule with world leading precision,
precise measurements of various two-body decays related to the extraction of angle $\phi_{2}$,
joining forces with Belle sample to offer most up-to-date information on $\phi_{3}$ from GLW and GLS analyses, and a
novel neutral charm tagger that nearly doubles the tagged $D$ meson sample size.

\section{Acknowledgement}

The author thanks to the Infosys Foundation for providing the leading edge travel grant.


\section*{References}

\begin{thebibliography}{99}
\bibitem{sumrule} M. Gronau, Phys. Lett. B \textbf{627}, 82 (2005).
\bibitem{flavortagger} F. Abudin\'en \textit{et al}. (Belle II Collaboration), Eur. Phys. J. C {\bf 82}, 283 (2022).
\bibitem{alpha}
M. Gronau and D. London, Phys. Rev. Lett. $\textbf{65}$,  3381 (1990).
\bibitem{GLW1} M. Gronau and D. London,
Phys. Lett. B, \textbf{253(3)}, 483–488( 1991).
\bibitem{GLW2}  M. Gronau and D. Wyler, Phys. Lett. B, \textbf{265(1)}, 172–176 (1991).
\bibitem{GLS} Z. Ligeti Y. Grossman and A. Soffer,
Phys. Rev. D, \textbf{67}, 071301 (2003).
\end{thebibliography}
\end{document}